# Understanding Subjectivity through the Lens of Motivational Context in Model-Generated Image Satisfaction

**Senjuti Dutta**
University of Tennessee, Knoxville
Knoxville, TN
sdutta6@vols.utk.edu

**Sherol Chen**
Google
Mountain View, CA
sherol@google.com

**Sunny Mak**
Google
Mountain View, CA
smak@google.com

**Amnah Ahmad**
Google
Mountain View, CA
amnahahmad@google.com

**Katherine M Collins**
Google Deepmind
Cambridge, UK
katiecollins@google.com

**Alena Butryna**
Google
Mountain View, CA
alenab@google.com

**Deepak Ramachandran**
Google
Mountain View, CA
ramachandrand@google.com

**Krishnamurthy Dj Dvijotham**
Google
Mountain View, CA
dvij@google.com

**Ellie Pavlick**
Google
Mountain View, CA
epavlick@google.com

**Ravi Rajakumar**
Google
Mountain View, CA
ravirajakumar@google.com

February 27, 2024

## Abstract

Image generation models are poised to become ubiquitous in a range of applications. These models are often fine-tuned and evaluated using human quality judgments that assume a universal standard, failing to consider the subjectivity of such tasks. To investigate how to quantify subjectivity, and the scale of its impact, we measure how assessments differ among human annotators across different use cases. Simulating the effects of ordinarily latent elements of annotators subjectivity, we contrive a set of motivations (t-shirt graphics, presentation visuals, and phone background images) to contextualize a set of crowdsourcing tasks. Our results show that human evaluations of images vary within individual contexts and across combinations of contexts. Three key factors affecting this subjectivity are image appearance, image alignment with text, and representation of objects mentioned in the text. Our study highlights the importance of taking individual users and contexts into account, both when building and evaluating generative models.

## 1 Introduction

Generative artificial intelligence (AI) models are set to have increased presence in numerous applications such as text [aut(2023), Chowdhery et al.(2022), Taylor et al.(2022), Touvron et al.(2023)], images [Ramesh et al.(2022), ALPACA(2023), Liu et al.(2023), Diffusion(2023), Midjourney(2023)], video [Ho et al.(2022), AI(2023c), Villegas et al.(2023), Runway(2023)], coding [GitHub(2023a), GitHub(2023b), Amazon(2023), Li et al.(2023), AI(2023b), CodeComplete(2023), Zheng et al.(2023)] and many other fields [Anyword(2023), CSM(2023), AI(2023a), Absci(2023)]. Human annotators providing feedback is a significant part of training these foundational AI models. Also, there's growing interest in exploring the use of human annotation to fine-tune those models via reinforcement learning with human feedback (RLHF). This annotation procedure inherently presumes a universal standard for what constitutes a "good" or "bad" output [Fisac et al.(2020)]. However, there are substantial grounds to question this assumption of a universal standard output, as human judgments are likely to exhibit considerable variation [Gordon et al.(2022)]. This subjectivity in opinion can be attributed not only to the specific task for which the AI system is being employed [Fan et al.(2022)] but also to the individual backgrounds [Aroyo and Welty(2015), Sen et al.(2015), Joshi et al.(2016)], personal identity [Geiger et al.(2020), Scheuerman et al.(2021)], and inherent biases [Wich et al.(2020), Miłkowski et al.(2021)].

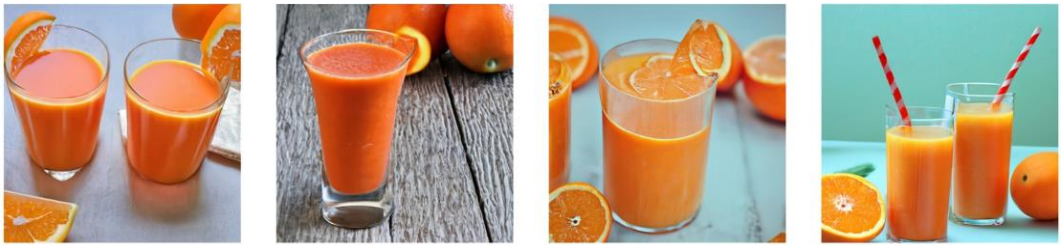

Figure 1: An example of AI-generated images for prompt "orange juice"

Our paper, therefore, investigates the assumption that there's a one-size-fits-all criterion for assessing the quality of AI-generated content, by directly measuring how annotations vary across annotators and use cases, which we refer to in this paper as *"motivational contexts"*. We focus specifically on image generation models, and we hone in on three different motivational contexts for which people may use these models: to create graphics for t-shirts, visuals for presentations, and background images for phones. Our goal is to understand the nuances in how people judge these AI-generated images (see Figure 1) from different motivational contexts.

To that end, we formulate the following research questions:

1. RQ1: Does the motivational context significantly affect the evaluation of the Image Generation Model output?
    (a) Does the evaluation for Image Generation model output differ significantly within the same motivational context?
    (b) Does the evaluation for Image Generation model output differ significantly across different motivational contexts?
2. RQ2: If significant differences are observed in RQ1, what are the sources contributing to these variations in the evaluation?

We find intriguing trends in variations in human evaluations, both within each motivational context and when comparing combinations of the aforementioned three motivational contexts. Moreover, our research points out that there are three key elements affecting subjectivity: how good the image looks, how well the image matches the text, and how well the image represents the object mentioned in the text. These findings echo humans are inherently subjective [Goyal et al.(2022)], a universal issue that extends beyond our specific domain. This inherent subjectivity presents a pervasive challenge across fields from healthcare to information retrieval, making our approach to understanding the role of motivational context in human evaluations of images particularly relevant and timely. Recognizing and accounting for this subjectivity is crucial for both creating and evaluating a more effective, inclusive, and adaptable generative model.

## 2 Related Work

To position our research, we first present a brief summary of the related work in two areas: metrics for measuring annotator (dis)agreement, and previous work regarding image annotation.

### 2.1 Metrics for measuring annotators agreement/ disagreement

Previous studies have presented various metrics for measuring agreement among annotators. In the context of crowdsourcing-based approaches, inter-rater reliability (IRR) is commonly utilized as a measure to assess the level of agreement when collecting annotated data [Hallgren(2012), Vargas et al.(2016)]. The IRR metrics encompass a wide range of coefficients, accommodating different experimental scenarios such as varying numbers of anno-tators, rating scales, agreement definitions, different types of crowdsourcing tasks, and assumptions regarding annotator interchangeability. Examples of these metrics include Scott's pi scott1955reliability, Cohen's kappa cohen1960coefficient, Siegel and Castellans kappa siegel1957nonparametric, Fleiss's kappa fleiss1971measuring, Byrt et al.'s kappa byrt1993bias, Krippendorff's alpha [Krippendorff(2011)].

While these traditional IRR metrics offer valuable insights for assessing annotator agreement, they may not fully account for the influence of varying motivational contexts. Therefore, in our study, we employ an adapted version of Krippendorff's alpha tailored to quantify agreement among annotators within and across motivational contexts.

### 2.2 Disagreement for Image Annotation

Prior studies have examined the evaluation of image datasets, revealing instances of disagreement among annotators. For instance, shevade2007modeling demonstrates the pervasive lack of consensus among annotators in these datasets, such as the Corel dataset, and in real-world personal image collections. However, the paper also highlights that while users may have low agreement with the entire group, they may have significantly higher agreement with a subset of the social network. In addition to the broader issue of inter-user semantic disagreement, previous research has specifically highlighted the challenge of inconsistent annotations within object identification in images. For example, rodrigues2017learning employed the LabelMe dataset [Russell et al.(2008)] to showcase that even when presented with a "gold standard" label for image categories, annotators often diverge in their classifications. These disagreements are not confined to laypeople but extend to expert annotators as well. Such variations underscore the innate subjectivity in image annotation, a challenge recognized across different domains from astronomical image classification [Smyth et al.(1994)] to medical imaging [Raykar et al.(2010)] and various others [Sharmanska et al.(2016), Rodrigues and Pereira(2018), Firman et al.(2018)]. Similar disagreements have been found when capturing uncertainty between annotators over both natural and synthetic images [Sucholutsky et al.(2023), Collins et al.(2023a), Collins et al.(2023b), Peterson et al.(2019), Sanders et al.(2022), Collins et al.(2022), Uma et al.(2022)].

Prior research primarily addresses the reliability and subjectivity of annotations but often overlooks how these factors may be influenced by the motivational context, particularly in relation to end-user goals. This represents a significant gap in our collective understanding, and our research aims to fill this gap by considering motivational contexts during the human evaluation of AI-generated images.

## 3 Method: Crowdsourcing Task

The crowdsourcing task was structured to assess whether the evaluation data for the image creation system is influenced by the context of end users' motivation. The process commences by inquiring about each annotator's satisfaction with the images generated by the model, along with given prompts, within each motivational context. Subsequently, having collected their satisfaction ratings for all three motivational contexts, each annotator is prompted to provide feedback on their holistic experience with the system through open-ended responses (see task instruction details in Appendix A). The data collection protocol underwent ethical review board approval. We focus on the evaluation task in this paper.

### 3.1 Evaluation Task

Our research question investigates understanding if the motivational context behind using AI-generated images influences the end user's annotations of those images. Although the ideal method would entail individualized observation of each user to ascertain their frame of reference for interacting with each image, logistical constraints make this impractical. To circumvent this limitation, we use motivational context as a proxy for user intent in image ratings. Specifically, the task began by asking annotators to evaluate a set of four [1] model-generated images for a given prompt on a 1-4 scale where 1 represents very poorly and 4 represents very well (see Figure 2) for the following three motivational contexts:

- Illustrative image for your Presentation: In this setting, annotators are tasked with evaluating each image to determine its fit for serving as a presentation image in a slide deck.
- Background image/ wallpaper for your Phone Screen: In this particular scenario, the annotator's task is to evaluate each image by considering if they would like to use it as a background for their phone or as phone wallpaper.
- Graphic for your T-shirt: In this specific context, the task for each annotator is to gauge the suitability of each image to use as a graphic for a t-shirt.

---

[1] We selected 4 model-generated images to reduce the cognitive load of annotators, which leads to more accurate and thoughtful evaluations.

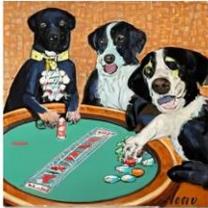

Figure 2: An example screen seen by annotators in our study; annotators are presented with a motivational context and a set of four model-generated images for the prompt

Each annotator was instructed explicitly to rate each image only based on the suitability of the image for each motivational context instead of focusing on the dimensions of the image as well as the frequency of the object mentioned in the image. For example, when annotators are presented with a set of four images relating to the prompt "orange juice," they need to evaluate each image based on its potential to be used for the allocated motivational context.

### 3.2 Data

We use the PartiPrompt set as our source of prompts. [2] This dataset, which is freely accessible to the public, comprises an extensive collection of more than 1600 English prompts. These prompts span a wide array of categories such as Abstract Concepts, Animals, Artifacts, Arts, Food and Beverages, Illustrations, Indoor and Outdoor Scenes, People, Products and Plants, Vehicles, and World Knowledge, making it appropriately diverse for the goals of our study. The original collection contains multiple variations of similar prompts, which could have introduced redundancy into our experiment. To tackle this issue, we eliminated such redundancies by randomly selecting one prompt from each group of similar prompts for inclusion in our study. For instance, multiple prompts existed around the theme of giant cobras like "A giant cobra snake made from corn, A giant cobra snake made from sushi, A giant cobra snake made from pancake, A giant cobra snake made from salad"; we opted for a random selection from this group rather than incorporating all similar variations. To that end, in our study, we chose a subset of 1338 prompts from the full PartiPrompts set.

### 3.3 Image Generation Models

To generate the images to be evaluated by annotators, we used a text-to-image diffusion model similar to DALL-E[Ramesh et al.(2021)] and Stable Diffusion [Rombach et al.(2022)]. The resulting images were automatically filtered for potentially harmful imagery.

### 3.4 Annotator Recruitment

We recruited 9 annotators from an internal crowdsourcing platform and put them into three groups randomly, with three annotators in each group. We chose 9 annotators because the total number needed to be a multiple of the number of motivational contexts we were studying. Due to time limitations, we opted for this specific number of annotators. On average, each annotator took three to four days to finish the task.

---

[2]https://github.com/google-research/parti/blob/main/PartiPrompts.tsv

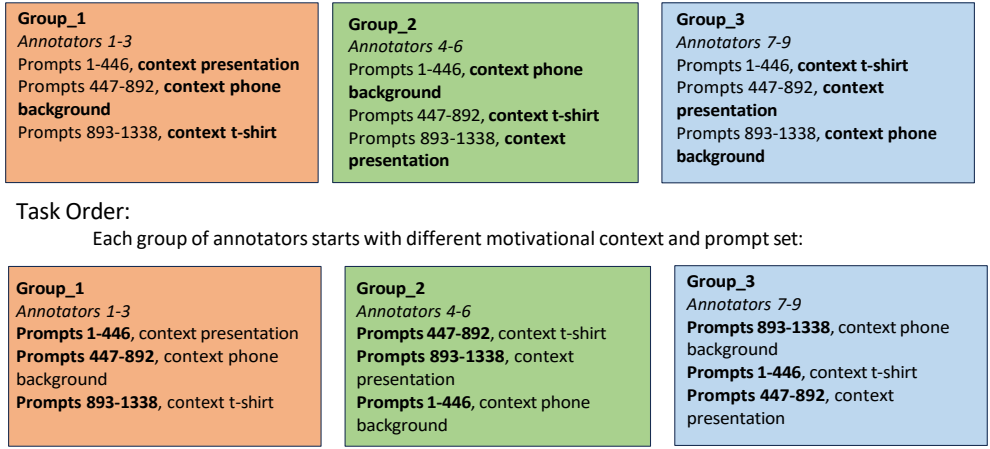

Figure 3: Experimental Design: Task Order

### 3.5 Experimental Design

Annotators were assigned to one of three groups (i.e., 3 annotators per group). Each group saw 1338 prompts and considered subsets of such prompts for each of the motivational contexts. For example, annotator group 1 starts their first set of questions with presentation as the motivational context, then sees a second (disjoint) set of questions with phone background as the motivational context, and then a last (again disjoint) set of questions with t-shirts as the motivational context. For each of annotator groups 2 and 3, the motivational context order is different and each annotator labels 1338 images in total. This design is shown in Figure 3.

### 3.6 Analysis

All evaluation questions were analyzed quantitatively except the open-ended responses. We first conducted a quantitative analysis of the outcomes related to each motivational context, followed by another quantitative analysis across every pair of motivational contexts, using inter-rater reliability (IRR) as a metric. Specifically, we use Krippendorffs alpha [Krippendorff(2011)] as it is well suited for the type of annotations utilized in this paper, namely, categorical data for image annotation. In this study, we adopt the Landis-Koch approach [Landis and Koch(1977)] to refer to the absolute interpretation of IRR. During our analysis, we focused solely on prompts evaluated using a 1-4 scale to examine the impact of motivational context on these specific ratings.

We calculate the within-context agreement using adapted IRR which is intended to capture the degree to which annotators tend to agree with the consensus opinion. Specifically, we take the IRR between two lists, $r_i$ and $r_{rest}$, where $r_i$ is the ratings of a given annotator, and $r_{rest}$ is the average ratings of the other annotators not including annotator $i$. See Appendix B for a complete description of our algorithm.

In order to better understand the influence of different motivational contexts on annotators' evaluation of images, we calculate the across-context agreement using adapted IRR for all possible pairs of contexts. This adapted metric allows us to compute a singular aggregated annotation for each motivational context, thereby facilitating a more straightforward comparative analysis. Specifically, we take the IRR between two lists, $r_{ctx_i}$ and $r_{ctx_j}$, where $r_{ctx_i}$ is the average rating for each image, across all annotators, within one context (e.g., phone background) and $r_{ctx_j}$ is the average rating for each image, across all annotators, within a different context (e.g., t-Shirt). See Appendix B for a complete description of our algorithm.

Additionally, we qualitatively analyze the open-ended responses provided by annotators in the task; i.e., their rationales for their context-conditioned ratings. We used a bottom-up approach to create codes in order to find themes [Braun and Clarke(2012)]. Our underlying intent is to provide an unbiased mechanism for capturing a wealth of sources to naturally observe how annotators disagree with each other not only within context but also across different motivational contexts.

| Motivational Context | IRR |
|---|---|
| Phone Background | 0.347 |
| Presentation | 0.436 |
| T-shirt | 0.447 |

Table 1: Calculated IRR value within each motivational context for three motivational contexts:(1) Phone Background Images, (2) Presentation Visuals, and (3) T-shirt

| Motivational Context Pair | IRR |
|---|---|
| T-shirt and Phone Background | 0.293 |
| Phone Background and Presentation | 0.302 |
| Presentation and T-shirt | 0.458 |

Table 2: IRR value across all possible pairs of three motivational contexts:(1)T-shirt and Phone Background, (2) Phone Background and Presentation, and (3) Presentation and T-shirt

## 4 Results

In this section, we showcase findings that address our research questions, focusing on whether the motivational setting influences how AI-generated images are evaluated both within a single motivational context and across varying motivational context scenarios. Additionally, we further delve into the factors underlying our observed variations.

### 4.1 Within Motivational Context

Table 1 shows the IRRs within each context. In the instance of the *phone background* motivational context, we observe the lowest IRR among all evaluated contexts. This might suggest a higher degree of subjectivity within this specific context, indicating that individual interpretations, biases, or personal tastes may have a more pronounced impact on the rating process. Conversely, the *t-shirt* motivational context exhibits the highest IRR value among the assessed contexts (see Table 1). The implication of this finding is that this motivational context may be less susceptible to subjectivity. For motivational context *presentation*, the calculated IRR is observed to be situated between the phone background and t-shirt contexts. These are interesting intuitions that could be confirmed by follow-up studies. Figure 4 shows an example of an image (prompt "a hedgehog") that has the highest agreement in the t-shirt motivational context but the lowest agreement for context phone background, whereas agreement for presentation lies in between these two motivational contexts.

### 4.2 Across Motivational Context

Table 2 shows the across-context agreement for all possible pairs of motivational contexts: (1) Phone Background and Presentation (2) Presentation and T-shirt (3) T-shirt and Phone Background. We see substantial variation in IRR depending on the motivational contexts being compared. Specifically, *t-shirt and phone background* motivational contexts exhibited the lowest IRR, implying that the types of images that one typically prefers as a phone background are different than the types of images that one prefers to put on a t-shirt. In contrast, the pair consisting of the *presentation and t-shirt* motivational contexts show the highest agreement among the annotators as shown in Figure 4. The IRR value for the pair including *phone background and presentation* is observed to be intermediate among all studied pairs of motivational contexts.

These findings augment our analysis of IRR within each motivational context. The *phone background* motivational context demonstrates the highest IRR when examined individually, suggesting it is the most subjective of the three contexts. In contrast, the *t-shirt* motivational context exhibits the lowest IRR, indicating it is the least subjective. The *presentation* motivational context falls in between these extremes. As a consequence, when considering pairs of motivational contexts, the phone background and t-shirt pair exhibit the highest IRR, indicating the most subjectivity. Conversely, the presentation and t-shirt pair demonstrates the lowest IRR, suggesting the least subjectivity across the evaluated context pairs as shown in Figure 4. These results highlight the importance of a more granular view of human evaluations of AI-generated images; human preferences for text-image alignment can be highly *and differentially* influenced by motivational context (i.e. annotators' frame of reference).

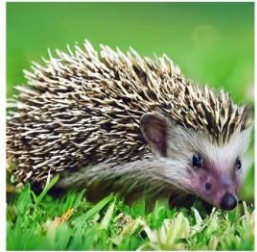

Prompt = "a hedgehog"

| Motivational Context | Annotation |
|---|---|
| Phone Background | [1,3,2] |
| Presentation | [3,2,4] |
| T-shirt | [4,3,3] |

| Across Motivational Context | Difference shown in terms of average |
|---|---|
| Phone Background and Presentation | Phone Background = 2<br>Presentation = 3 |
| Presentation and T-shirt | Presentation = 3<br>T-shirt = 3.33 |
| T-shirt and Phone Background | T-shirt = 3.33<br>Phone Background = 2 |

Figure 4: Example annotation within each motivational context and average of annotation shown across each possible pair of motivational context from three contexts

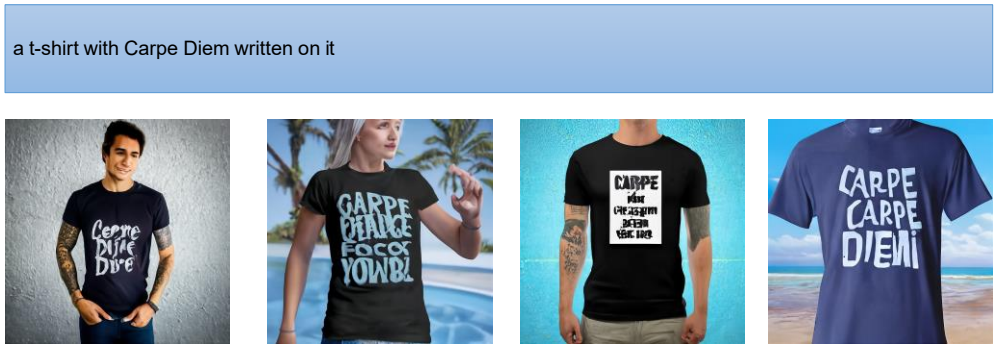

a t-shirt with Carpe Diem written on it

Figure 5: Model-generated images for a specific prompt "a t-shirt with Carpe Diem written on it"

## 4.3 Takeaways from Quantitative Analysis

Overall, our analyses indicate a distinct trend: the *phone background* motivational context is especially subjective, both alone and when paired with other motivational contexts. These initial findings suggest room for individual biases in this setting, but further research is needed to either confirm or challenge our initial observations.

## 4.4 Sources behind Subjectivity

Given the above trends, we perform a qualitative analysis of the sources of subjectivity among annotators for image annotation for three motivational contexts. The purpose of this approach is to provide intuition and better understand the factors contributing to subjectivity in the evaluation process. We found the following three themes from the coding:

- Image Quality: This theme includes factors such as resolution, clarity, color balance, composition, and the presence of noise or artifacts in the image. For example, for the prompt "a t-shirt with Carpe Diem written on it" (model-generated images for this prompt are shown in Fig 5) for motivational context *t-shirt* one annotator mentioned that image quality is distorted and therefore provided a rating of 2.
- Image alignment with prompt: This theme refers to the degree to which an image corresponds, matches, or adheres to a given prompt. It measures the relevancy and appropriateness of an image in relation to a predefined directive or task. For the same example, for prompt "a t-shirt with Carpe Diem written on it" for motivational context *t-shirt* other annotator provided a rating of 3 mentioning that:

    *"The image aligns with the prompt "(P1)*

- Representation and existence of an object in the image: This theme pertains to the presence and portrayal of distinct objects or text within an image. It assesses both the existence of these elements and the clarity, legibility, and appropriateness of their depiction, contributing to the overall interpretation and understanding of the image. For the same example, for prompt "a t-shirt with Carpe Diem written on it" for motivational context *t-shirt* the other annotator provided a rating of 1 mentioning that:

    *"T-shirt lacks "Carpe Diem""(P3)*

These themes suggest that individuals' perceptions and evaluations of images are influenced by multiple, often intertwined factors. "Image Quality" refers to the technical and aesthetic properties of an image; "Image and Prompt Alignment" points to the contextual relevance of an image in relation to a specific prompt and "Representation and Existence of Object/Text in Image" concerns the visibility and interpretation of specific elements within the image. The interplay of at least two themes or all three themes simultaneously can contribute to the diversity in perspectives and, hence, subjective assessments. Overall, these themes underscore the multifaceted nature of image interpretation and evaluation. They highlight that subjectivity in image assessments may arise from differences in individuals' aesthetic preferences, their interpretation of the prompt, and their ability to identify and interpret specific elements within the image.

## 5 Limitations

Although our work has taken steps to understand subjectivity through the lens of motivational context in model-generated image satisfaction, it is important to acknowledge the limitations of our work.

Firstly, our study does not examine the scenario in which no motivational context is provided. For a more comprehensive understanding of the influence of motivational context, we recommend that future research undertake a comparative analysis of annotator agreement across annotated images, contrasting those evaluated with a motivational context against those assessed without any contextual cues. Such a comparison could offer valuable insights into the differential impacts of motivational context on the evaluation process. Additionally, our study is confined to only three distinct motivational contexts. For a more comprehensive understanding, we encourage future studies to encompass a wider array of such contexts. Doing so would shed light on whether the subjectivity we observed is consistent across a more extensive range of situations and reveal how these variations might manifest differently based on different motivational contexts. Our annotator pool was relatively small, consisting of only 9 individuals, primarily due to our budget constraints. Subsequent research would benefit from considering a larger and potentially more diverse annotators sample. A more extensive dataset would enable a more robust exploration of how human evaluations may vary. Finally, in our analysis, we focused solely on prompts evaluated using a 1-4 scale to examine the impact of motivational context on these specific ratings. For future research, we recommend expanding the scope to include all rating categories, as well as instances where annotators indicated that the prompt was not contextually appropriate. This could offer additional insights into how often prompts and what types of prompts are skipped or disregarded across different motivational contexts.

## 6 Discussion

Our study aimed to understand the effects of motivational context on the subjective nature of AI-generated image annotations by human annotators, focusing specifically on three distinct motivational contexts: presentation images, phone background images, and graphics for t-shirts. Our results reveal a significant level of variance in annotations even within the same motivational context, thereby challenging the assumption that human annotations serve as a monolithic ground truth. These findings suggest some important directions for future work in machine learning, particularly in tasks reliant on human-annotated data.

*Contextual Variability as a Critical Feature for Model Training* The observed variance in human annotations across different motivational contexts suggests an immediate requirement for future ML models to be context-aware (adaptive to the environment). The introduction of context as an auxiliary feature could enable ML algorithms to dynamically adapt user behavior, rendering them more suitable for specific use cases or applications.

*Uncertainty Modeling and Ensemble Methods toward Personalized Outputs* Our results point to the importance of considering subjectivity in the human appraisal of AI-generated output. Given that humans may differ in their judgments of the quality of output for their context, our work hints at the importance of considering uncertainty in model training and evaluation: for both the reward and foundational generative model. Different mechanisms (e.g.ensemble techniques [3]) may be employed to enable the model to navigate the landscape of diverse human opinions, rather than

---

[3] https://www.sciencedirect.com/topics/computer-science/ensemble-method

A PREPRINT - FEBRUARY 27, 2024A PREPRINT - FEBRUARY 27, 2024converge to the "average" user response [Gordon et al.(2022)]. This could make systems more robust and individually tailored to users.

*Data Augmentation Strategies and Annotator Reliability* Our study's findings on the variable reliability of human annotators across contexts offer a potential strategy for data augmentation. Future models may benefit from weighting annotations based on the consistency or reliability of the annotator, thus incorporating an additional layer of complexity and richness into the training data.

*Extending to Multimodal models* Beyond using these results for image-based models, they could also be incorporated into multimodal models [Danny Driess(2023), Huang et al.(2023)]. The context-specific insights generated from our research could be particularly valuable for these more complex models, potentially offering a more holistic understanding of how different types of data can be integrated effectively to account for human subjectivity.

*Overall*, our study points to an intricate relationship between motivational context and human subjectivity in image annotations, offering both challenges and avenues for future research. As machine learning models increasingly integrate human-generated data, acknowledging and adapting to this subjectivity becomes imperative for improving model reliability, generalizability, and personalization. Our findings thus provide a step towards developing ML models that can nimbly navigate the complexity of human subjectivity across diverse motivational contexts. Future work will benefit from leveraging these insights, driving both theoretical advancement and practical applications of image annotation and interpretation and beyond that.

## 7 Conclusion

In the ever-evolving landscape of generative AI models, evaluation and fine-tuning often hinges on human evaluations of the produced output, an approach that implicitly assumes a universal standard of quality for good or bad output. Our research investigates this assumption by examining the variations in human judgments for image evaluation across three different motivational contexts: t-shirt graphics, presentation visuals, and phone background images. We find significant variability in human judgments not only within each motivational context but also across each possible pair of motivational contexts. Notably, the phone background context shows a heightened level of subjectivity, both on its own and in combination with other contexts. Conversely, evaluations for motivational context t-shirt appears to be the least subjective. Evaluations for motivational context presentation fall in between the context phone background and t-shirt. In addition to that, our result shows three key factors contributing to this subjectivity: aesthetic quality, text-image alignment, and how well the image represents the object mentioned in the text. Our results underscore that the motivational context and hence the annotator's frame of reference significantly affects the human evaluation of the image generation model output. This indicates a caution against the one-size-fits-all approach to AI evaluation and underscores the importance of taking individual users and contexts into account, both when building and evaluating generative AI models.

## References

[aut(2023)] 2023. *GPT-4 Technical Report*. Technical Report.

[Absci(2023)] Absci. 2023. *Biologics Drug Discovery Absci*. https://www.absci.com/ Accessed: 17-May-2023.

[AI(2023a)] A. AI. 2023a. *AIVA - The AI composing emotional soundtrack music*. https://www.aiva.ai/ Accessed: 17-May-2023.

[AI(2023b)] B. AI. 2023b. *BLACKBOX AI*. https://www.useblackbox.io/ Accessed: 17-May-2023. [AI(2023c)]

M. AI. 2023c. *Make-A-Video by Meta AI*. https://makeavideo.studio/ Accessed: 17-May-2023. [ALPACA(2023)]

ALPACA. 2023. *Alpaca - Humans AI Models for Image Generation*. https://www.getalpaca.io/ Accessed: 17-May-2023.

[Amazon(2023)] Amazon. 2023. *Compañero de codificación ML - Amazon Code-Whisperer - Amazon Web Services*. https://aws.amazon.com/es/codewhisperer/ Accessed: 17-May-2023.

[Anyword(2023)] Anyword. 2023. *Anyword Copy Intelligence and Generative AI Built for Marketers*. https://anyword.com/ Accessed: 17-May-2023.

[Aroyo and Welty(2015)] Lora Aroyo and Chris Welty. 2015. Truth is a lie: Crowd truth and the seven myths of human annotation. *AI Magazine* 36, 1 (2015), 15–24.

[Braun and Clarke(2012)] Virginia Braun and Victoria Clarke. 2012. Thematic analysis. (2012).
9

[Chowdhery et al.(2022)] Aakanksha Chowdhery, Sharan Narang, Jacob Devlin, Maarten Bosma, Gaurav Mishra, Adam Roberts, Paul Barham, Hyung Won Chung, Charles Sutton, Sebastian Gehrmann, et al. 2022. Palm: Scaling language modeling with pathways. *arXiv preprint arXiv:2204.02311* (2022).

[CodeComplete(2023)] CodeComplete. 2023. *CodeComplete: AI Coding Assistant for Enterprise*. https://codecomplete.ai/ Accessed: 17-May-2023.

[Collins et al.(2023a)] Katherine Maeve Collins, Matthew Barker, Mateo Espinosa Zarlenga, Naveen Raman, Umang Bhatt, Mateja Jamnik, Ilia Sucholutsky, Adrian Weller, and Krishnamurthy Dvijotham. 2023a. Human Uncertainty in Concept-Based AI Systems. In *Proceedings of the 2023 AAAI/ACM Conference on AI, Ethics, and Society*. 869–889.

[Collins et al.(2023b)] Katherine M Collins, Umang Bhatt, Weiyang Liu, Vihari Piratla, Ilia Sucholutsky, Bradley Love, and Adrian Weller. 2023b. Human-in-the-loop mixup. In *Uncertainty in Artificial Intelligence*. PMLR, 454–464.

[Collins et al.(2022)] Katherine M Collins, Umang Bhatt, and Adrian Weller. 2022. Eliciting and learning with soft labels from every annotator. In *Proceedings of the AAAI Conference on Human Computation and Crowdsourcing*, Vol. 10. 40–52.

[CSM(2023)] CSM. 2023. *CSM AI - 3D World Models*. https://csm.ai/ Accessed: 17-May-2023.

[Danny Driess(2023)] Pete Florence Danny Driess. 2023. *Palm-E: Embodied Multimodal Language*. Google Research Blog. https://blog.research.google/2023/03/palm-e-embodied-multimodal-language.html Accessed: 2023-05-17.

[Diffusion(2023)] S. Diffusion. 2023. *Stable Diffusion Online*. https://stablediffusionweb.com/ Accessed: 17-May-2023.

[Fan et al.(2022)] Mingming Fan, Xianyou Yang, TszTung Yu, Q Vera Liao, and Jian Zhao. 2022. Human-ai collaboration for ux evaluation: Effects of explanation and synchronization. *Proceedings of the ACM on Human-Computer Interaction* 6, CSCW1 (2022), 1–32.

[Firman et al.(2018)] Michael Firman, Neill DF Campbell, Lourdes Agapito, and Gabriel J Brostow. 2018. Diversenet: When one right answer is not enough. In *Proceedings of the IEEE Conference on Computer Vision and Pattern Recognition*. 5598–5607.

[Fisac et al.(2020)] Jaime F Fisac, Monica A Gates, Jessica B Hamrick, Chang Liu, Dylan Hadfield-Menell, Malayandi Palaniappan, Dhruv Malik, S Shankar Sastry, Thomas L Griffiths, and Anca D Dragan. 2020. Pragmatic-pedagogic value alignment. In *Robotics Research: The 18th International Symposium ISRR*. Springer, 49–57.

[Geiger et al.(2020)] R Stuart Geiger, Kevin Yu, Yanlai Yang, Mindy Dai, Jie Qiu, Rebekah Tang, and Jenny Huang. 2020. Garbage in, garbage out? Do machine learning application papers in social computing report where human-labeled training data comes from?. In *Proceedings of the 2020 Conference on Fairness, Accountability, and Transparency*. 325–336.

[GitHub(2023a)] GitHub. 2023a. *About GitHub Copilot for Individuals - GitHub Docs*. https://docs.github.com/en/copilot/overview-of-github-copilot/about-github-copilot-for-individual Accessed: 17-May-2023.

[GitHub(2023b)] GitHub. 2023b. *GitHub Copilot X: The future of AI-powered software development*. https://www.linkedin.com/pulse/github-copilot-x-future-ai-powered-software-development-github/?tr Accessed: 17-May-2023.

[Gordon et al.(2022)] Mitchell L Gordon, Michelle S Lam, Joon Sung Park, Kayur Patel, Jeff Hancock, Tatsunori Hashimoto, and Michael S Bernstein. 2022. Jury learning: Integrating dissenting voices into machine learning models. In *Proceedings of the 2022 CHI Conference on Human Factors in Computing Systems*. 1–19.

[Goyal et al.(2022)] Nitesh Goyal, Ian D Kivlichan, Rachel Rosen, and Lucy Vasserman. 2022. Is your toxicity my toxicity? exploring the impact of rater identity on toxicity annotation. *Proceedings of the ACM on Human-Computer Interaction* 6, CSCW2 (2022), 1–28.

[Hallgren(2012)] Kevin A Hallgren. 2012. Computing inter-rater reliability for observational data: an overview and tutorial. *Tutorials in quantitative methods for psychology* 8, 1 (2012), 23.

[Ho et al.(2022)] Jonathan Ho, William Chan, Chitwan Saharia, Jay Whang, Ruiqi Gao, Alexey Gritsenko, Diederik P Kingma, Ben Poole, Mohammad Norouzi, David J Fleet, et al. 2022. Imagen video: High definition video generation with diffusion models. *arXiv preprint arXiv:2210.02303* (2022).

[Huang et al.(2023)] Shaohan Huang, Li Dong, Wenhui Wang, Yaru Hao, Saksham Singhal, Shuming Ma, Tengchao Lv, Lei Cui, Owais Khan Mohammed, Qiang Liu, et al. 2023. Language is not all you need: Aligning perception with language models. *arXiv preprint arXiv:2302.14045* (2023).

[Joshi et al.(2016)] Aditya Joshi, Pushpak Bhattacharyya, Mark Carman, Jaya Saraswati, and Rajita Shukla. 2016. How do cultural differences impact the quality of sarcasm annotation?: A case study of indian annotators and american text. In *Proceedings of the 10th SIGHUM Workshop on Language Technology for Cultural Heritage, Social Sciences, and Humanities*. 95–99.

[Krippendorff(2011)] Klaus Krippendorff. 2011. Computing Krippendorff's alpha-reliability. (2011).

[Landis and Koch(1977)] J Richard Landis and Gary G Koch. 1977. The measurement of observer agreement for categorical data. *biometrics* (1977), 159–174.

[Li et al.(2023)] Y. Li, D. Choi, J. Chung, N. Kushman, J. Schrittwieser, R. Leblond, T. Eccles, J. Keeling, F. Gimeno, A. D. Lago, T. Hubert, P. Choy, C. de Masson dAutume, I. Babuschkin, X. Chen, P.-S. Huang, J. Welbl, S. Gowal, A. Cherepanov, J. Molloy, D. J. Mankowitz, E. S. Robson, P. Kohli, N. de Freitas, K. Kavukcuoglu, and O. Vinyals. 2023. *AlphaCode*. https://alphacode.deepmind.com/ Accessed: 17-May-2023.

[Liu et al.(2023)] Guan-Horng Liu, Arash Vahdat, De-An Huang, Evangelos A Theodorou, Weili Nie, and Anima Anandkumar. 2023. IΘ2 SB: Image-to-Image Schr\" odinger Bridge. *arXiv preprint arXiv:2302.05872* (2023).

[Midjourney(2023)] Midjourney. 2023. *Midjourney*. https://www.midjourney.com/home/?callbackUrl=%2Fapp%2F Accessed: 17-May-2023.

[Miłkowski et al.(2021)] Piotr Miłkowski, Marcin Gruza, Kamil Kanclerz, Przemyslaw Kazienko, Damian Grimling, and Jan Kocoń. 2021. Personal bias in prediction of emotions elicited by textual opinions. In *Proceedings of the 59th Annual Meeting of the Association for Computational Linguistics and the 11th International Joint Conference on Natural Language Processing: Student Research Workshop*. 248–259.

[Peterson et al.(2019)] Joshua C Peterson, Ruairidh M Battleday, Thomas L Griffiths, and Olga Russakovsky. 2019. Human uncertainty makes classification more robust. In *Proceedings of the IEEE/CVF International Conference on Computer Vision*. 9617–9626.

[Ramesh et al.(2022)] Aditya Ramesh, Prafulla Dhariwal, Alex Nichol, Casey Chu, and Mark Chen. 2022. Hierarchical text-conditional image generation with clip latents. *arXiv preprint arXiv:2204.06125* 1, 2 (2022), 3.

[Ramesh et al.(2021)] Aditya Ramesh, Mikhail Pavlov, Gabriel Goh, Scott Gray, Chelsea Voss, Alec Radford, Mark Chen, and Ilya Sutskever. 2021. Zero-shot text-to-image generation. In *International Conference on Machine Learning*. PMLR, 8821–8831.

[Raykar et al.(2010)] Vikas C Raykar, Shipeng Yu, Linda H Zhao, Gerardo Hermosillo Valadez, Charles Florin, Luca Bogoni, and Linda Moy. 2010. Learning from crowds. *Journal of machine learning research* 11, 4 (2010).

[Rodrigues and Pereira(2018)] Filipe Rodrigues and Francisco Pereira. 2018. Deep learning from crowds. In *Proceedings of the AAAI conference on artificial intelligence*, Vol. 32.

[Rombach et al.(2022)] Robin Rombach, Andreas Blattmann, Dominik Lorenz, Patrick Esser, and Björn Ommer. 2022. High-resolution image synthesis with latent diffusion models. In *Proceedings of the IEEE/CVF conference on computer vision and pattern recognition*. 10684–10695.

[Runway(2023)] Runway. 2023. *Runway - Everything you need to make anything you want*. https://runwayml.com/ Accessed: 17-May-2023.

[Russell et al.(2008)] Bryan C Russell, Antonio Torralba, Kevin P Murphy, and William T Freeman. 2008. LabelMe: a database and web-based tool for image annotation. *International journal of computer vision* 77 (2008), 157–173.

[Sanders et al.(2022)] Kate Sanders, Reno Kriz, Anqi Liu, and Benjamin Van Durme. 2022. Ambiguous images with human judgments for robust visual event classification. *Advances in Neural Information Processing Systems* 35 (2022), 2637–2650.

[Scheuerman et al.(2021)] Morgan Klaus Scheuerman, Alex Hanna, and Emily Denton. 2021. Do datasets have politics? Disciplinary values in computer vision dataset development. *Proceedings of the ACM on Human-Computer Interaction* 5, CSCW2 (2021), 1–37.

[Sen et al.(2015)] Shilad Sen, Margaret E Giesel, Rebecca Gold, Benjamin Hillmann, Matt Lesicko, Samuel Naden, Jesse Russell, Zixiao Wang, and Brent Hecht. 2015. Turkers, scholars," arafat" and" peace" cultural communities and algorithmic gold standards. In *Proceedings of the 18th acm conference on computer supported cooperative work & social computing*. 826–838.

[Sharmanska et al.(2016)] Viktoriia Sharmanska, Daniel Hernández-Lobato, Jose Miguel Hernandez-Lobato, and Novi Quadrianto. 2016. Ambiguity helps: Classification with disagreements in crowdsourced annotations. In *Proceedings of the IEEE Conference on Computer Vision and Pattern Recognition*. 2194–2202.

[Smyth et al.(1994)] Padhraic Smyth, Usama Fayyad, Michael Burl, Pietro Perona, and Pierre Baldi. 1994. Inferring ground truth from subjective labelling of venus images. *Advances in neural information processing systems* 7 (1994).

[Sucholutsky et al.(2023)] Ilia Sucholutsky, Ruairidh M Battleday, Katherine M Collins, Raja Marjieh, Joshua Peterson, Pulkit Singh, Umang Bhatt, Nori Jacoby, Adrian Weller, and Thomas L Griffiths. 2023. On the informativeness of supervision signals. In *Uncertainty in Artificial Intelligence*. PMLR, 2036–2046.

[Taylor et al.(2022)] Ross Taylor, Marcin Kardas, Guillem Cucurull, Thomas Scialom, Anthony Hartshorn, Elvis Saravia, Andrew Poulton, Viktor Kerkez, and Robert Stojnic. 2022. Galactica: A large language model for science. *arXiv preprint arXiv:2211.09085* (2022).

[Touvron et al.(2023)] Hugo Touvron, Thibaut Lavril, Gautier Izacard, Xavier Martinet, Marie-Anne Lachaux, Timothée Lacroix, Baptiste Rozière, Naman Goyal, Eric Hambro, Faisal Azhar, et al. 2023. Llama: Open and efficient foundation language models. *arXiv preprint arXiv:2302.13971* (2023).

[Uma et al.(2022)] Alexandra Uma, Dina Almanea, and Massimo Poesio. 2022. Scaling and disagreements: Bias, noise, and ambiguity. *Frontiers in Artificial Intelligence* 5 (2022), 818451.

[Vargas et al.(2016)] Saúl Vargas, Richard McCreadie, Craig Macdonald, and Iadh Ounis. 2016. Comparing overall and targeted sentiments in social media during crises. In *Proceedings of the International AAAI Conference on Web and Social Media*, Vol. 10. 695–698.

[Villegas et al.(2023)] R. Villegas, M. Babaeizadeh, P.-J. Kindermans, H. Moraldo, H. Zhang, M. T. Saffar, S. Castro, J. Kunze, and D. Erh. 2023. *Phenaki*. https://phenaki.video/ Accessed: 17-May-2023.

[Wich et al.(2020)] Maximilian Wich, Hala Al Kuwatly, and Georg Groh. 2020. Investigating annotator bias with a graph-based approach. In *Proceedings of the fourth workshop on online abuse and harms*. 191–199.

[Zheng et al.(2023)] Q. Zheng, X. Xia, X. Zou, Y. Dong, S. Wang, Y. Xue, Z. Wang, L. Shen, A. Wang, Y. Li, T. Su, Z. Yang, and J. Tang. 2023. *GitHub - THUDM/CodeGeeX: CodeGeeX: An Open Multilingual Code Generation Model*. https://github.com/THUDM/CodeGeeX Accessed: 17-May-2023.

## A  Task Instruction to Annotators

In this task, you will be given a total of around 1.5K prompts across 3 motivational contexts. For each prompt your job is to do the following:

1. Read the task description carefully
2. For each motivational context, you will be given a set of prompts. For each given prompt the model will generate a set of 4 images.
   - If the prompt does not make any sense with the motivational context to you then check the does not match the context box and submit the question.
   - Otherwise for each prompt based on the Likert scale question choose the rating and provide your rationale for your selected rating.
3. Submit
4. Exit Survey: After completing all evaluation tasks, write your responses about the overall experience interacting with the system (optional)
5. Submit

## B  Algorithms Details

### B.1  Within-Context Algorithm

Our dataset is organized into a dictionary named images. In this dictionary, each image identifier(id) maps to a list of ratings provided for that image. For example, if images[img8] = [1, 2, 1], this indicates that the image with the identifier img8 has received three ratings: 1 from the first annotator, 2 from the second, and 1 from the third.

The algorithm proceeds as follows:

- Initialize Lists: We create two empty lists, *all_r0* and *all_rrest*, to store the ratings from the "first annotator" and the "average ratings from the rest of the two annotators," respectively, for each image.
- Iterate Over Images: We loop through each image identifier in the images dictionary.
- Representation of Rating: We extract the first annotator's rating (denoted as *r_0*) from the list of ratings for each image. We calculate the average rating from the rest of the annotators (denoted as *r_rest*) for each image.
- Append to Lists: We append *r_0* and *r_rest* to *all_r0* and *all_rrest* lists, respectively.
- Output: Finally, we calculate the inter-rater reliability between *all_r0* and *all_rrest* using Krippendorff's alpha. For a more robust measure, we execute the above algorithm in a "round-robin" fashion: first comparing *r_0* against the average of the rest, then *r_1* against the average of the rest excluding *r_1*, and so on. The final IRR is then the average of these individual IRR calculations.

## B.2 Across-Context Algorithm

Our dataset is represented by a nested dictionary, images, where each image ID is mapped to another dictionary. This second-level dictionary maps various contexts (e.g., "t-shirt," "presentation", "phone background") to a list of ratings for that image within that context.

For instance, images["t-shirt"][img8] = [1, 2, 1] means that within the context labeled "t-shirt," the image with ID img8 received ratings of 1, 2, and 1 from three different annotators, respectively. The algorithm proceeds as follows:

1. Initialization: Create two empty lists all_c0 and all_c1 that will hold the average ratings for each image under two contexts being compared.
2. Iterating Through Context Pairs: Loop through all unique pairs of contexts (context0, context1) from the predefined list of possible contexts, all_possible_pairs_of_contexts which are (t-shirt, phone background), (presentation, t-shirt) and (phone background, t-shirt).
3. Iterating Through Image IDs: For each unique image ID in images,
    - Calculate Average Ratings: Compute the average rating within the first context, denoted c_0, and the average rating within the second context, denoted c_1.
    - Append to Lists: Add the computed averages c_0 and c_1 to all_c0 and all_c1, respectively.
4. IRR Calculation: After exiting the loop, we compute the IRR between all_c0 and all_c1.
5. Output: Print the pair of contexts (context0, context1) along with the computed IRR value for that pair.